\documentclass[acmsmall,screen,nonacm]{acmart}

\usepackage{amsmath,amssymb,booktabs,tabularx}
\usepackage{graphicx}
\usepackage{listings}
\setcopyright{none}
\renewcommand\footnotetextcopyrightpermission[1]{}
\acmDOI{}
\acmISBN{}
\acmYear{2026}
\copyrightyear{2026}
\newcommand{\method}{Additive GRPO}
\newcommand{\oce}{\mathrm{OCE}}
\newcommand{\bts}{\mathrm{BTS}}
\newcommand{\ind}{\mathbf{1}}

\title[Oracle Conversion in Specification-Based Test Generation]{Correct Tests Are Not Enough: Measuring and Training Oracle Conversion in Specification-Based Test Generation}
\author{Yunhao Liang}
\affiliation{%
  \institution{Chengdu Institute of Computer Applications, Chinese Academy of Sciences}
  \country{China}}
\affiliation{%
  \institution{University of Chinese Academy of Sciences}
  \country{China}}

\author{Chengguang Gan}
\affiliation{%
  \institution{Independent Researcher}
  \country{Japan}}

\author{Ruixuan Ying}
\affiliation{%
  \institution{Institute of Multidisciplinary Research for Advanced Materials (IMRAM), Tohoku University}
  \country{Japan}}

\author{Hanjun Wei}
\affiliation{%
  \institution{University of Chinese Academy of Sciences}
  \country{China}}

\author{Zhe Cui}
\affiliation{%
  \institution{Chengdu Institute of Computer Applications, Chinese Academy of Sciences}
  \country{China}}
\affiliation{%
  \institution{University of Chinese Academy of Sciences}
  \country{China}}

\author{Shiwen Ni}
\affiliation{%
  \institution{Artificial Intelligence Research Institute,\linebreak[1] Shenzhen University of Advanced Technology}
  \country{China}}
\renewcommand{\shortauthors}{Author information to be added}
\begin{document}

\begin{abstract}
Generating tests from a natural-language specification requires both an input that exposes faulty behavior and a correct expected output. These requirements need not improve together: a model can increase test correctness by choosing easier inputs, or discover useful inputs whose expected outputs it cannot predict. We study this interaction through executable reward decomposition and suite-level oracle-conversion measurement. Our generator jointly emits five input--output tests in one response. During training, audited reference programs provide correctness feedback, while a fixed bank of faulty programs provides two utility signals: potential input kill and effective kill after checking the generated output. An additive GRPO objective preserves both signals without requiring execution at inference time. On an audited TC-Bench split with 506 training and 142 evaluation tasks, three independently trained Qwen3.5-9B runs at step 75 increase full-test correctness from 28.59\% to 42.54\%, input kill from 24.06\% to 25.27\%, and effective full kill from 12.23\% to 14.15\%. Matched 50-step ablations reveal a trade-off: removing kill rewards yields higher correctness and slightly higher full kill, but lowers input kill to 21.60\%. A fixed-input source--oracle crossover on 64 training-pool tasks attributes the principal NoKill-to-FullKill difference to harder input selection rather than worse output prediction on identical inputs. These results identify oracle conversion as a measurable bottleneck and show the benefits and limits of preserving input-utility feedback in joint test generation.
\end{abstract}
\ccsdesc[500]{Software and its engineering~Software testing and debugging}
\keywords{test generation, test oracles, executable feedback, reinforcement learning, fault detection}
\maketitle

\section{Introduction}
Tests written before an implementation must express intended behavior rather than merely reproduce the behavior of an existing program. This requirement makes specification-based test generation attractive for test-driven development: a developer supplies a programming requirement, and a model proposes inputs together with their expected outputs. The tests can then reject candidate implementations that violate the requirement. Unlike regression-test generation, this setting cannot assume that an implementation under test is already a trustworthy source of expected behavior. The distinction is an instance of the test-oracle problem~\cite{oracle}, and it remains important when a language model produces the entire test.

Automated testing has long separated input exploration from behavioral checking. Feedback-directed generation uses execution to identify useful input sequences~\cite{randoop}; whole-suite search and assertion generation use an implementation to construct regression tests~\cite{evosuite}. Neural oracle generation addresses the complementary problem of inferring assertions from program context~\cite{toga}. More recent systems combine language models with coverage feedback or execution-based filtering to improve tests for available code~\cite{coverup,testgen}. In contrast, our deployment interface contains a natural-language statement and its input/output conventions, but no candidate implementation, reference solution, public input/output example, or interactive execution feedback. The output is a small complete test suite, not test code that requires a focal-method signature.

This interface creates two distinct opportunities for failure. A test can contain an incorrect expected output, even when its input is executable. It can also be entirely correct yet fail to distinguish any faulty program in the evaluation bank. These failures call for different learning signals. Rewarding only correctness can favor input distributions on which the model is comfortable computing outputs. Rewarding discriminative inputs can instead favor cases whose outputs require more difficult reasoning. When fault-detection credit is awarded only after the generated output is correct, a useful input with an incorrect output receives no direct credit for its latent value.

The tension itself is not new. UTGen explicitly studies the need to generate revealing inputs and accurate outputs without access to a gold implementation, and uses generated tests in automated debugging~\cite{utgen}. UTRL trains a tester with adversarial feedback from a coder~\cite{utrl}, while CURE co-evolves coding and testing policies~\cite{cure}. These methods establish the importance of discriminative test training. They leave room for a narrower empirical question: \emph{when an executable reward changes the inputs chosen by a joint test generator, how much of the resulting fault-detection opportunity survives the model's own expected outputs?} A higher correctness rate alone cannot answer this question, and neither can an input-only kill score.

We address this question with \emph{oracle conversion}: the extent to which potentially detectable faulty programs are actually detected by fully correct generated tests. For each task, we execute a generated input against audited reference programs to obtain a consensus output. The same input defines a set of faulty programs that it could distinguish with that output. Only tests whose generated outputs match the consensus contribute to effective detection. Taking unions within a suite avoids awarding repeated credit for detecting the same faulty program. The difference between potential and effective kill then quantifies lost opportunities; their suite-unique aggregate ratio measures conversion efficiency.

This accounting motivates a simple joint training objective, \method{}. In addition to format, count, input-validity, and full-correctness rewards, it includes both effective full kill and potential input kill. The latter remains available when the input executes correctly but the model predicts the wrong output. All components are computed from programs during training. At inference, the trained policy emits five complete tests in a single response, without a selector or execution loop. The experimental focus is the allocation of executable reward and the behavior that it produces.

We evaluate Qwen3.5-9B on an audited TC-Bench subset~\cite{tcbench}. The principal comparison uses one frozen Base output set and three independent training seeds. At step 75, \method{} improves full correctness by 13.94 percentage points, potential input kill by 1.22 points, and effective full kill by 1.92 points. All three seeds improve these metrics relative to Base, although one seed's input-kill gain is only 0.02 points. The result is a joint improvement of modest utility magnitude, not a claim that reward decomposition eliminates the oracle problem.

The ablations explain why the distinction matters. At an equal 50-step training budget, NoKill reaches 47.04\% full correctness and 14.36\% full kill, both above Additive's 40.66\% and 13.67\%. Yet NoKill's input kill falls below Base in every seed. Additive preserves a higher mean input kill, but converts fewer of its opportunities. Thus, neither correctness nor input utility ranks the methods in the same way as effective detection. To separate input selection from oracle ability, we also cross three input-source policies with three output-prediction policies while holding the input bytes fixed. For the primary NoKill-to-FullKill contrast, output prediction improves on identical inputs, while the selected input distribution becomes harder. This rules out a simple account in which kill-oriented training merely damages all oracle prediction.

The paper makes three contributions:
\begin{itemize}
  \item A suite-unique oracle-conversion protocol that jointly reports test correctness, potential fault detection, effective fault detection, and the opportunities lost at the output stage.
  \item A matched executable-reward study of joint test generation, including NoKill, FullKill-only, and Additive objectives, three independent seeds, and an optimizer-preserving training-length continuation.
  \item A source--oracle crossover that separates the difficulty of chosen inputs from the ability to predict their outputs, a distinction hidden in aggregate test scores.
\end{itemize}

\section{Background and Related Work}
\label{sec:related}
\subsection{Input Generation, Oracles, and Fault Detection}
Execution-guided test generation is well established. Randoop builds sequences using feedback about legality, contracts, and redundancy~\cite{randoop}. EvoSuite optimizes whole suites and derives assertions that characterize observed behavior~\cite{evosuite}. These tools demonstrate why a useful test is more than an independently sampled input. Their principal interface, however, includes code to execute. Our deployment interface deliberately excludes such code, while retaining reference execution as a training and evaluation instrument.

An expected result must be justified by the specification, not by whichever program is available. The oracle literature distinguishes the challenges of constructing and checking such expectations~\cite{oracle}. TOGA learns exceptional and assertion oracles from contextual information, including settings with incomplete implementations~\cite{toga}. Our tests use stdin/stdout rather than API-level assertions, and the generator chooses the input as well as its output. This choice makes the oracle's input distribution endogenous to training. A model that appears more accurate may simply ask itself easier questions.

Fault-based testing measures whether tests distinguish faulty behavior~\cite{mutation}. We use the term \emph{kill} for consistency with the evaluation code, but our fault bank contains dataset-provided wrong solutions, not necessarily syntactic mutants produced by a mutation operator. A kill score therefore describes detection within a finite program bank. It is not a measure of all possible defects, and the number of wrong solutions is not the number of tests required to cover them. Several programs can fail on the same input; conversely, a single incorrect program can be difficult to distinguish with any of the five chosen inputs.

\subsection{Language Models for Test Construction}
Implementation-aware LLM testing systems can exploit code and a test infrastructure. CoverUp guides regression-test generation using coverage and execution feedback~\cite{coverup}, while TestGen-LLM accepts proposed improvements after checking successful building, reliable passing, and added coverage~\cite{testgen}. These workflows address test improvement for available implementations. Predicting expected outputs from a requirement alone removes that source of behavioral evidence.

Execution also supports code generation and repair. Self-Debugging uses generated code and execution behavior to revise programs~\cite{selfdebug}; AlphaCode combines large-scale generation with filtering and clustering for programming problems~\cite{alphacode}. Our study concerns the tests that such workflows might use, rather than a downstream code-generation system. We consequently do not infer improved program synthesis or developer productivity from a higher test-bank kill score.

Test quality affects the reliability of code evaluation. APPS~\cite{apps} and TACO~\cite{taco} provide programming tasks from competitions, while EvalPlus shows that augmenting tests can reveal errors missed by an original suite~\cite{evalplus}. HardTests studies the construction of stronger executable tests for coding models~\cite{hardtests}. TC-Bench instead supports analysis of relationships between tests and collections of correct and wrong programs through a binary-matrix perspective~\cite{tcbench}. We use its program banks to make potential and effective detection separately observable. Our audited subset is a new evaluation protocol over that source data, not a replacement benchmark with independently established oracle guarantees.

\subsection{Learning Test Quality from Feedback}
Reinforcement learning has been used to improve static test-quality metrics and reduce undesirable test patterns~\cite{rlutf}. UTGen targets revealing tests for automated debugging and explicitly recognizes the joint difficulty of selecting inputs and predicting outputs~\cite{utgen}. UTRL obtains discrimination feedback from coder samples and trains the tester adversarially~\cite{utrl}. CURE similarly studies co-evolution between coding and testing policies~\cite{cure}. Our setting differs in using a fixed, audited bank of wrong programs and holding the training backbone, rollout budget, and inference interface constant across nested reward objectives.

Recent work also studies how suite utility should be assigned within a rollout. Ockhamareto combines a Pareto-gated bonus with test-segment credit for mutation effectiveness and suite concision~\cite{ockhamareto}. We retain response-level credit and a requested suite size of five. The contribution is therefore not the first use of mutation-style rewards, additive quality rewards, or fine-grained test utility. It is the measurement and experimental separation of potential input utility from its conversion through a model-generated oracle in specification-only generation.

We use GRPO~\cite{grpo}, building on the clipped policy-update family of PPO~\cite{ppo}, with LoRA adaptation~\cite{lora}. The implementation uses VERL/HybridFlow~\cite{hybridflow} and vLLM~\cite{vllm}. These are infrastructure choices, not proposed algorithmic contributions. Likewise, chain-of-thought prompting can change the reasoning budget~\cite{cot}, but the matched experiments disable thinking mode and request only the completed JSON suite. Their gains cannot be attributed to adding a second reasoning phase at deployment.

\section{Task and Oracle-Conversion Measures}
\label{sec:measures}
\subsection{Specification-Only Complete-Test Generation}
A task $q$ consists of a natural-language problem statement, input and output conventions, and constraints. The generator receives this specification and returns one suite
\begin{equation}
 S_q=\{(x_i,\hat y_i)\}_{i=1}^{m_q},\qquad m_q\leq B,\quad B=5,
\end{equation}
after deterministic parsing, input deduplication, and truncation to the evaluation budget. Each pair describes one independent stdin/stdout run. The requested JSON response has this shape:
\begin{lstlisting}
{"tests":[{"input":"...","output":"..."}]}
\end{lstlisting}
A missing slot is not replaced by another model call. We distinguish the requested budget from the number recovered from the response so that incomplete or repetitive suites cannot benefit from a smaller denominator.

The evaluator, but not the generation prompt, has access to an audited reference bank $\mathcal R_q$ and a faulty-program bank $\mathcal W_q$. Let $V_q(x)$ indicate that the reference executions satisfy the configured success and agreement checks. When $V_q(x)=1$, their normalized consensus output is $y_q^*(x)$. We call such an input \emph{reference-valid}. This is an operational criterion: agreement of the reference programs does not prove that every natural-language constraint holds for $x$.

For a generated pair, define full correctness as
\begin{equation}
 c_q(x,\hat y)=V_q(x)\ind[\operatorname{norm}(\hat y)=y_q^*(x)].
\end{equation}
Normalization standardizes line endings and strips surrounding and line-trailing whitespace. It does not turn arbitrary output text into a semantically equivalent answer. Problems known to require special judges or permit multiple valid outputs are excluded from this protocol.

\subsection{Potential and Effective Detection}
For a reference-valid input $x$, let $D_q(x)\subseteq\mathcal W_q$ be the faulty programs whose executions fail or whose normalized outputs disagree with $y_q^*(x)$. For an invalid input, set $D_q(x)=\varnothing$. The suite's potential and effective detection sets are
\begin{align}
 P_q&=\bigcup_{(x,\hat y)\in S_q}D_q(x),\\
 E_q&=\bigcup_{(x,\hat y)\in S_q:\,c_q(x,\hat y)=1}D_q(x).
\end{align}
The associated rates are
\begin{equation}
 O_q=\frac{|P_q|}{|\mathcal W_q|},\qquad
 K_q=\frac{|E_q|}{|\mathcal W_q|}.
 \label{eq:kill}
\end{equation}
We refer to $O_q$ as \emph{input kill} or \emph{potential kill}, and to $K_q$ as \emph{full kill} or \emph{effective kill}. By construction, $E_q\subseteq P_q$ and $K_q\leq O_q$. An incorrect generated output is not allowed to reject programs: otherwise, a false assertion could manufacture an apparent kill. Potential kill instead asks what the same input would detect if its expected output were supplied correctly.

The unit of detection is a task--program pair. If two tests reject the same faulty program, that program contributes once to the suite union. If one has an incorrect output but the other is fully correct, the opportunity is still converted. This distinction is important: summing per-test kill rates can overcount redundant detections, and multiplying an average correctness rate by an average kill rate does not reproduce the suite union.

Consider an illustrative suite whose inputs could detect faulty programs $\{a,b\}$ and $\{b,c\}$. If only the first expected output is correct, the suite has three potential detections and two effective detections. The lost opportunity is $c$, not every program rejected by the second input. This example describes the accounting rule, not an observed benchmark result.

\subsection{Aggregate Scores and Their Denominators}
For $N$ evaluation tasks, input validity and full correctness use the fixed slot budget:
\begin{equation}
 V=\frac{\sum_q\sum_{i=1}^{m_q}V_q(x_i)}{NB},\qquad
 C=\frac{\sum_q\sum_{i=1}^{m_q}c_q(x_i,\hat y_i)}{NB}.
\end{equation}
We report $O=N^{-1}\sum_qO_q$ and $K=N^{-1}\sum_qK_q$, assigning zero to failed tasks. This task-macro average gives a problem with five wrong programs the same weight as one with nineteen. Task success denotes successful recovery and evaluation of at least one test, not a guarantee of a complete or correct suite. We separately report the proportion of tasks for which all five slots are fully correct.

Oracle-conversion efficiency uses the unique detection sets:
\begin{equation}
 \oce=\frac{\sum_q|E_q|}{\sum_q|P_q|}.
 \label{eq:oce}
\end{equation}
It is undefined if there are no potential detections. Equation~\ref{eq:oce} weights opportunities rather than tasks. The reported multi-seed OCE is the arithmetic mean of each run's ratio, not a ratio formed after averaging task-macro kill rates. Accordingly, $K=O\cdot\oce$ is not generally an identity for the reported aggregate columns. The macro conversion gap, $G=O-K$, is the complementary absolute measure of unrealized potential.

OCE alone can favor a conservative generator that finds very few opportunities and converts all of them. We therefore report it alongside $C$, $O$, and $K$. As a compact joint diagnostic, we also use the task-level harmonic mean
\begin{equation}
 \bts=\frac{1}{N}\sum_q
 \begin{cases}
  \dfrac{2C_qO_q}{C_q+O_q},&C_q+O_q>0,\\[3pt]
  0,&\text{otherwise},
 \end{cases}
 \qquad C_q=\frac{1}{B}\sum_i c_q(x_i,\hat y_i).
\end{equation}
We call this the Balanced TDD Score (BTS). It discourages a high value on only one of correctness and potential detection, but it does not replace effective full kill: the correct tests need not be the tests that distinguish faulty programs. BTS uses a standard harmonic mean to summarize the two requirements. Its name denotes the intended use of the tests, not a measured downstream TDD benefit.

\subsection{Input-Kill Bins}
To examine conversion within the generated distribution, each reference-valid input receives the per-test potential rate $d_i=|D_q(x_i)|/|\mathcal W_q|$. We group inputs into five bins: $0$, $(0,0.10]$, $(0.10,0.25]$, $(0.25,0.50]$, and $(0.50,1]$. Within a bin we report oracle correctness and mean effective per-test kill $c_q(x_i,\hat y_i)d_i$, together with the number of valid slots. These curves exclude missing and invalid inputs because their potential difficulty is undefined. They do not use the suite-union aggregation of Equation~\ref{eq:kill}. Nor do they hold inputs fixed across methods: they describe where each policy spends its generation budget. Section~\ref{sec:crossover} supplies the separate fixed-input experiment needed to examine input selection and oracle ability.

\section{Executable Reward Decomposition}
\label{sec:method}
\subsection{One Generator, Two Utility Signals}
The policy $\pi_\theta$ jointly generates the input and output strings. Reference programs are external evaluators during training, not a second model in the deployed generator. For each sampled suite, the evaluator first determines which inputs are reference-valid and obtains their consensus outputs. It then evaluates the wrong-program bank twice conceptually: potential utility uses every reference-valid input, whereas effective utility uses only fully correct pairs. The same executions can be reused for both sets. This preserves the input's utility signal without pretending that an incorrect test is safe to use.

The distinction matters under a small suite budget. Suppose one rollout selects easy inputs and predicts all outputs correctly, while another finds a revealing input but miscomputes its output. A correctness-only objective favors the former. An effective-kill objective may also fail to distinguish the latter from a rollout that finds no revealing input at all. The potential-kill term supplies that missing distinction. The full-correctness and effective-kill terms still reward completing the test correctly. This is a learning objective for competing requirements, not a guarantee that every policy update improves both.

\subsection{Reward Definition}
Let $F$ be a binary strict-format indicator, $Q$ a count-and-uniqueness score, $V_q$ the fraction of the five slots that are reference-valid, and $C_q$ the fraction that are fully correct. For a strict-format response, the training reward is
\begin{align}
 r_{\mathrm{add}}(q,S)={}&0.05F+0.05Q+0.15V_q+0.25C_q\nonumber\\
 &+0.25K_q+0.25O_q-0.10L,
 \label{eq:reward}
\end{align}
where $L$ is an excess-length penalty. Format requires a JSON object with exactly the key \texttt{tests}, a list value, and exactly two string fields, \texttt{input} and \texttt{output}, in every item. A strict-format failure receives $-1-0.10L$. Format validity is distinct from having five valid or fully correct tests; those conditions do not form additional hard gates in this objective.

If the response contains $n$ raw tests and the parser recovers $m\leq5$ nonempty unique inputs, the count component is
\begin{equation}
 Q=\max\!\left(0,1-\frac{|n-5|}{5}\right)
 \min\!\left(\frac{m}{n},1\right),
\end{equation}
with $Q=0$ when $n=0$. Duplicate inputs therefore consume generation budget without increasing execution coverage. The length penalty is
\begin{equation}
 L=\min\!\left(1,\frac{\max(0,\ell-16000)}{16000}\right),
\end{equation}
where $\ell$ is the response length in characters. Executable scoring excludes individual inputs beyond 8,000 characters, while the generation backend separately enforces a response-token limit.

The format component is constant among strict-format responses. It is not claimed to supply within-group ranking when every response has the correct schema. We retain it, together with the count term, to match the implemented quality objective across the ablations. In contrast, count, validity, correctness, and both kill components can vary among well-formed responses.

\begin{table}[t]
\caption{Nested reward objectives. All arms share the format-failure rule, count, validity, correctness, and length settings. Removed coefficients are not redistributed. The names denote reward ablations, not reproductions of other published systems.}
\label{tab:rewards}
\centering
\begin{tabular}{lccc}
\toprule
Objective & Quality part & Effective kill $K_q$ & Potential kill $O_q$\\
\midrule
NoKill & $0.05F+0.05Q+0.15V_q+0.25C_q-0.10L$ & 0 & 0\\
FullKill-only & Same & 0.25 & 0\\
Additive & Same & 0.25 & 0.25\\
\bottomrule
\end{tabular}
\end{table}

Table~\ref{tab:rewards} defines the matched objectives. NoKill tests whether improvements can arise from basic quality feedback without discriminative reward. FullKill-only adds detection credit available through correct generated outputs. Additive adds potential detection independent of output correctness. Because the coefficients of the remaining terms are unchanged, these are operational ablations of the implemented reward. They do not independently identify every effect of reward scale, reward covariance, and group normalization.

\subsection{Group-Relative Optimization}
For each prompt, we sample $G=8$ complete suites from the current rollout policy. GRPO~\cite{grpo} uses each scalar reward relative to the other rewards in that prompt's group:
\begin{equation}
 A_i=\frac{r_i-\overline r}{\operatorname{std}(r_1,\ldots,r_G)+\epsilon}.
\end{equation}
The implementation applies the usual clipped likelihood-ratio policy loss, aggregated over response tokens, with a reference-policy KL penalty. This follows the PPO family of conservative updates~\cite{ppo}; it does not train a value critic. Every response token receives the suite-level advantage. We do not assign separate input and output token advantages in the main method.

Group-relative normalization changes the optimization interpretation of an additive reward. A term contributes only insofar as it differentiates sampled suites, and its effect depends on covariance with the other terms. If all eight rewards are identical, the centered task reward supplies no relative advantage, although a KL regularizer may still contribute to the update. Adding $O_q$ can distinguish useful-but-incorrect tests, but cannot help a group in which all inputs are invalid or all suites cover the same program set. We do not filter such groups or replace them using a reward-dependent resampling rule in these experiments.

LoRA~\cite{lora} adapts the policy with rank 32 and scaling parameter 64. VERL/HybridFlow~\cite{hybridflow} orchestrates policy updates and vLLM rollout generation~\cite{vllm}. The base model remains the same across reward objectives. There is no supervised warm-start stage in the principal experiment; prior SFT and decoupled explorations are not folded into its reported training budget.

\subsection{Execution and Deployment Boundary}
Correct and wrong C++ programs are precompiled and cached. At reward time, execution uses per-task time limits with the configured timeout allowance, output limits, and process isolation. At least five reference programs must support the consensus; retained references must execute successfully and agree. A timeout or disagreement invalidates the reference oracle for that input rather than turning it into negative evidence against a wrong program. Wrong-program failures count toward detection only after a usable reference oracle is available.

For $B$ inputs, $R$ reference programs, and $W$ faulty programs, scoring requires at most $B(R+W)$ program executions before cache reuse. Compilation is amortized across steps, and potential/effective kill share the resulting execution matrix. The expensive component is execution feedback during training and offline evaluation, not a learned critic. At deployment, the model makes a single joint generation call, and the reference and wrong-program banks are unavailable to it. The returned tests are model predictions; the method does not certify them for arbitrary unseen requirements.

\section{Experimental Design}
\label{sec:setup}
We organize the evaluation around four questions. \textbf{RQ1:} Does executable additive training improve correctness and effective detection together relative to the same base model? \textbf{RQ2:} What does each kill-reward component change at a matched training budget? \textbf{RQ3:} How much potential detection is lost through generated outputs, and is the loss explained by input selection or oracle prediction? \textbf{RQ4:} How stable are the results across seeds and additional training steps?

\subsection{Audited Data and Program Banks}
Our source is TC-Bench~\cite{tcbench}. We audit the programs and judging requirements before using them for executable rewards. The cleaning process removes known multiple-output or special-judge tasks, excludes tasks with exact duplicate wrong solutions, deduplicates identical correct solutions, and checks reference programs against official sample behavior and available dynamic audit inputs. Problems without enough retained executable references are excluded. These operations address observed problems in the data; passing them does not constitute a proof of program correctness.

The resulting Quality-v2 split contains 704 tasks, partitioned into 506 training, 56 validation, and 142 evaluation tasks with rank stratification. Table~\ref{tab:data} summarizes the retained banks. Their sizes are moderate enough to execute the full wrong-program bank rather than sampling a different set each reward call. All main metrics use the audited banks associated with this split. Scores from older 150- or 162-task protocols are not mixed directly into the main comparison.

\begin{table}[t]
\caption{Audited TC-Bench Quality-v2 data. The rank and wrong-program-count ranges coincide in this retained subset; neither is an intrinsic oracle-reasoning difficulty measure.}
\label{tab:data}
\centering
\begin{tabular}{lrrrrr}
\toprule
Split & Tasks & Rank range & Mean rank & References/task & Wrong/task\\
\midrule
Train & 506 & 5--19 & 9.76 & 5--8 & 5--19\\
Validation & 56 & 5--19 & 9.79 & 5--8 & 5--19\\
Evaluation & 142 & 5--19 & 9.74 & 5--8 & 5--19\\
\bottomrule
\end{tabular}
\end{table}

The prompt builder removes recognized example sections from the description and does not append the dataset's sample-input or sample-output fields. Correct and wrong solutions are never inserted into generation prompts. The same joint-test template is used for the matched Base and trained policies. Data partitioning is by task, not by individual test case. The 142 tasks were nevertheless examined during earlier project experiments and during training-length selection. We therefore treat them as a fixed project evaluation set, not as a previously untouched estimate for an adaptively selected method.

\subsection{Training and Evaluation Settings}
The matched backbone is Qwen3.5-9B. Each reward arm starts independently from Base using seeds 101, 102, and 103. Training uses four NVIDIA A800 80GB GPUs, four prompts per update, eight rollouts per prompt, a policy minibatch of four prompts, and a per-GPU microbatch of one. Rollout temperature is 0.8 and top-$p$ is 1.0. The learning rate is $10^{-5}$, the configured warm-up fraction is 0.03, and the KL-loss coefficient is 0.001. Thinking mode is disabled. We use gradient checkpointing, LoRA on linear modules, and token-mean loss aggregation. The maximum prompt and response lengths are 12,288 and 4,096 tokens, respectively.

The equal-budget ablation stops at 50 updates. The length experiment continues Additive to step 75 while restoring model, optimizer, learning-rate scheduler, data-loader, and random-number-generator state, and retaining the strict-schema reward. It is an exact continuation, not an additional adapter-only restart with a new optimizer. A historical optimizer-reset branch exists in the project but is not the step-75 result reported here. Fifty and seventy-five updates correspond to approximately 200 and 300 prompt presentations at batch four, about 0.40 and 0.59 passes over 506 tasks. These step counts should not be interpreted as epochs.

Evaluation requests five complete tests at temperature zero, with generation seed 42 and thinking disabled for the matched models. A task is a stateless model call: previous tasks are not carried into its context. The evaluator retains at most five unique normalized inputs. It can extract a syntactically valid JSON value from a response and supports the existing parser's field aliases; it does not repair malformed JSON or synthesize missing outputs. This parsing behavior is shared by all reported joint-generation results. It is broader than the strict schema rewarded during training, so task success is not labeled strict-format accuracy.

Completed model failures remain in the output records. They are not repeatedly sampled until they succeed. Failed requests, unparseable responses, and missing slots contribute zero under the fixed denominators in Section~\ref{sec:measures}. A suite with only four usable tests can still receive credit for those tests, but its fifth slot remains zero. An exact-five condition is not retroactively applied as a whole-task evaluation gate.

\subsection{Comparisons and Statistical Reporting}
The controlled comparisons are Base and the three reward objectives in Table~\ref{tab:rewards}. We evaluate the deterministic Base once and compare every training seed against that same output set. Trained arms are reported as three-seed arithmetic means, accompanied by individual values, percentage-point changes, and directions of change. A seed is the unit of training replication; the five test slots within a task are dependent observations.

We also retain joint-generation results from larger local models and API endpoints as capability references. These use the same task set and five-test scoring budget after rejudgment, but differ in model size, serving implementation, and potentially internal reasoning behavior. In particular, the GPT-labeled endpoint was accessed through a third-party compatible service, whose underlying model identity we cannot independently verify. These rows are labeled by their recorded endpoint names and are not equal-compute competitors or evidence about an official provider release. They contextualize the task, rather than isolate the effect of our reward.

\section{Results}
\label{sec:results}
\subsection{RQ1: Joint Improvement Relative to Base}
Table~\ref{tab:main} reports the controlled comparison. Base produces fully correct tests in 28.59\% of the five-slot budget and detects 12.23\% of wrong programs per task. Its inputs could detect 24.06\% if their outputs were correct. The 11.82-point gap is a substantial loss of already-discovered opportunity, rather than evidence that the model never reaches useful inputs.

\begin{table}[t]
\caption{Matched Qwen3.5-9B results on 142 tasks. All numbers are percentages; trained rows are three-seed means. $V$: reference-valid slots; $C$: fully correct slots; $O$: input kill; $K$: full kill. OCE is opportunity-weighted within each run and then averaged across seeds. Bold identifies the largest mean among the equal-budget 50-step arms only.}
\label{tab:main}
\centering\small
\setlength{\tabcolsep}{4.5pt}
\begin{tabular}{lrrrrrrrr}
\toprule
Method & Success & $V$ & $C$ & $O$ & $K$ & OCE & BTS & All-five\\
\midrule
Base & 91.55 & 76.20 & 28.59 & 24.06 & 12.23 & 49.70 & 17.74 & 6.34\\
\midrule
NoKill-50 & \textbf{99.77} & 80.56 & \textbf{47.04} & 21.60 & \textbf{14.36} & \textbf{65.89} & 22.73 & \textbf{14.32}\\
FullKill-only-50 & 99.06 & 80.89 & 45.59 & 23.59 & 14.19 & 59.48 & \textbf{23.74} & 10.56\\
Additive-50 & 99.06 & \textbf{81.50} & 40.66 & \textbf{24.43} & 13.67 & 55.73 & 21.74 & 7.04\\
\midrule
Additive-75 & 97.89 & 79.81 & 42.54 & 25.27 & 14.15 & 56.51 & 23.27 & 7.28\\
\bottomrule
\end{tabular}
\end{table}

At step 75, Additive reaches 42.54\% full correctness, 25.27\% input kill, and 14.15\% full kill. Relative to Base, the changes are $+13.94$, $+1.22$, and $+1.92$ points, respectively, calculated from unrounded values. Input validity also rises by 3.62 points, and BTS rises by 5.53 points. The gain in full kill is approximately 15.7\% relative to Base's 12.23\%, but its absolute magnitude remains below two percentage points. Reporting both views avoids mistaking a relative improvement for a large absolute increase in detected faults.

All three step-75 seeds improve $C$, $O$, $K$, and BTS relative to Base. The input-kill margins are uneven: seed 101 gains only 0.02 points, while seed 103 gains 2.96 points. Additive therefore provides a replicated direction of joint improvement on this project split, not a uniformly large input-utility improvement. Moreover, only 7.28\% of tasks have all five tests fully correct. The slot-level correctness gain does not mean that most returned suites are ready to use without review.

OCE rises from 49.70\% to 56.51\%, indicating that the trained model converts a larger fraction of its discovered opportunities. It also discovers more opportunities, leaving an absolute macro gap of 11.12 points. Thus, both potential and effective detection can improve while the gap changes little. The quantities answer different questions: how much is discovered, what fraction is converted, and how much is ultimately detected.

\subsection{RQ2: Reward Components Change the Chosen Trade-off}
All three 50-step objectives substantially improve full correctness over Base. NoKill leads the matched arms in both correctness (47.04\%) and mean full kill (14.36\%). Basic validity and correctness feedback alone can therefore improve effective detection in this setting.

The cost is a reduction in potential detection. NoKill's mean input kill is 21.60\%, down 2.46 points from Base, and the reduction occurs in all three seeds. FullKill-only recovers input kill to 23.59\%, but remains below Base in each seed. Additive reaches 24.43\%, the highest mean at the matched budget, while its full correctness and OCE are lower. This pattern is consistent with the reward exposing utility in input regions whose outputs are more difficult to predict.

The effective-kill results do not monotonically follow the added reward components. FullKill-only and Additive have means of 14.19\% and 13.67\%, respectively, below NoKill's 14.36\%. Additive-50 is the only arm whose mean improves all four central metrics $C$, $O$, $K$, and BTS over Base, but it is not the best arm for full kill or BTS. The result supports retaining potential utility when that is a requirement; it does not establish that the chosen weights maximize the final deployment objective.

This distinction also explains why a single composite score is insufficient. FullKill-only has the highest 50-step BTS, whereas NoKill has the highest full kill and Additive the highest input kill. A suite can receive a good balance score even when its correct outputs are concentrated on its less useful inputs. Effective union-based kill must remain visible alongside any correctness--utility summary.

\subsection{RQ3: Conversion Loss and Input Difficulty}
\label{sec:conversion}
\begin{figure}[t]
\centering
\includegraphics[width=\linewidth]{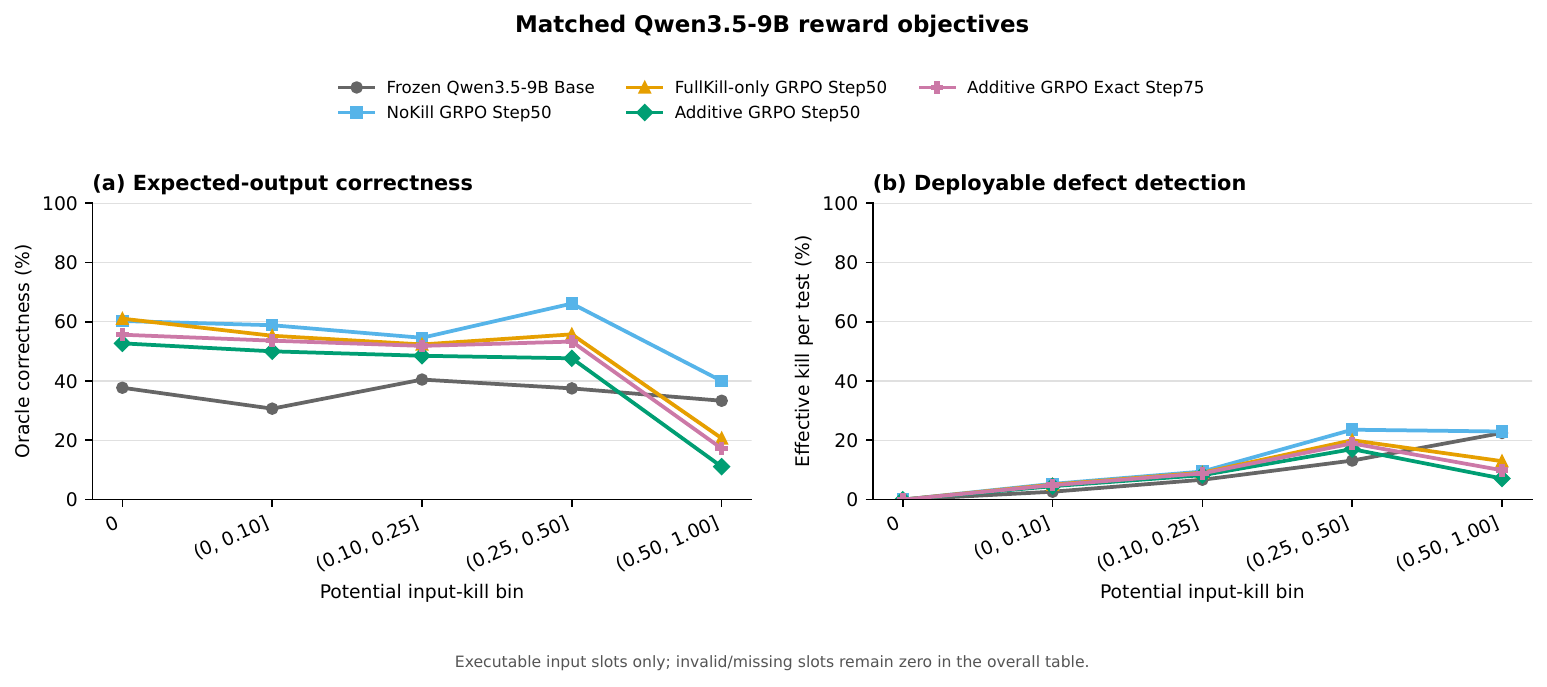}
\caption{Oracle conversion across generated-input kill bins on the 142-task evaluation set. Trained-policy slots are pooled across three seeds for the curves. Full kill here is an effective \emph{per-test} rate, not the suite union in Table~\ref{tab:main}. Bins contain different inputs across methods, and the high-kill bin is sparse; the figure is descriptive rather than a fixed-input causal comparison.}
\label{fig:conversion}
\end{figure}

Figure~\ref{fig:conversion} traces oracle correctness and effective per-test kill across input-kill bins. The curves are not universally monotone: NoKill, for example, is more accurate in the $(0.25,0.50]$ bin than in the preceding bin. They nevertheless expose a marked failure in the highest-potential region for the kill-trained 9B policies. Among Additive-50's 27 valid slots with per-input kill above 0.5, only three have correct outputs, an oracle accuracy of 11.11\%. FullKill-only has 20.69\% and NoKill 40.00\% in the corresponding bins. Their counts are 29 and 25, respectively.

\begin{table}[t]
\caption{Highest-potential bin, $d_i>0.5$, from the same offline records as Figure~\ref{fig:conversion}. Counts are valid input slots, pooled across the three runs for trained arms.}
\label{tab:highbin}
\centering
\begin{tabular}{lrrrr}
\toprule
Method & Slots & Mean input kill & Oracle correct & Effective per-test kill\\
\midrule
Base & 6 & 72.70 & 33.33 & 22.49\\
NoKill-50 & 25 & 65.51 & 40.00 & 22.96\\
FullKill-only-50 & 29 & 67.63 & 20.69 & 12.96\\
Additive-50 & 27 & 67.78 & 11.11 & 7.11\\
Additive-75 & 29 & 68.68 & 17.24 & 9.90\\
\bottomrule
\end{tabular}
\end{table}

The suite-level opportunity audit leads to a compatible observation. Relative to FullKill-only, the step-50 Additive runs newly reach 204 task--program opportunities in the recorded paired attribution. Only 36.76\% of these newly reached opportunities are converted through correct outputs. Relative to NoKill, 254 new opportunities are reached and 32.28\% are converted. These are counts from matched program-set differences, not additional independent tasks. They show why reaching more faulty programs need not increase the effective union: new opportunities can be lost at the output stage while previously effective detections are also lost.

\subsubsection{Fixed-input source--oracle crossover}
\label{sec:crossover}
The bin analysis cannot determine whether a policy has become intrinsically worse at output prediction, because it evaluates that policy's own inputs. We therefore run a separate crossed diagnostic on 64 tasks drawn from the training pool. For each seed, NoKill, FullKill-only, and Additive first generate their ordinary joint suites. We discard the source outputs, freeze the input bytes and order, and ask each of the three policies to fill all outputs in one joint JSON scaffold. This produces nine source--target cells per seed and 27 cells overall. The experiment is diagnostic on training-pool tasks; it is not an additional held-out generalization result.

Table~\ref{tab:crossover} reports output accuracy on source-valid slots, counting target failures as zero. Changing the target within a row changes the oracle policy on the same source inputs. Changing the source within a column changes the selected input distribution while keeping the target policy fixed. Because some targets fail strict schema or input-copy checks, the paired oracle contrast additionally restricts comparison to identical slots on common successful tasks. For the NoKill-to-FullKill contrast, the pooled common support is 561 source--task instances and 2,385 valid slots across seeds and source policies. All 27 cells complete without request failures.

\begin{table}[t]
\caption{Fixed-input crossover on 64 training-pool tasks. Cells are E2E output accuracy on source-valid inputs, averaged across three seeds. The rightmost column is the source's potential input kill, which is unchanged by the target oracle policy.}
\label{tab:crossover}
\centering
\begin{tabular}{lrrrr}
\toprule
Input source & NoKill oracle & FullKill oracle & Additive oracle & Input kill\\
\midrule
NoKill & 62.71 & 64.07 & 63.33 & 18.91\\
FullKill-only & 59.64 & 59.17 & 59.88 & 20.92\\
Additive & 56.23 & 57.99 & 55.76 & 20.67\\
\bottomrule
\end{tabular}
\end{table}

For the primary NoKill-to-FullKill contrast, the source input kill increases by 2.02 points. Averaging the source change across target policies yields a 3.81-point oracle-accuracy cost, positive in every seed. In contrast, changing the target from NoKill to FullKill on identical common-support inputs \emph{improves} oracle accuracy by 1.18 points, also in every seed. Strict input/schema preservation changes by $-0.35$ points and E2E joint completion by $+0.88$ points, with mixed seed directions. The main observed loss is therefore associated with which inputs are selected, not a general deterioration of the oracle on unchanged inputs.

The secondary FullKill-to-Additive contrast is less favorable. Source input kill changes by $-0.25$ points on this diagnostic pool, source-associated oracle cost is $+2.90$ points, and matched target oracle accuracy changes by $-0.83$ points. Potential-kill reward does not uniformly improve input selection in every task subset, and it can coexist with a smaller conditional-oracle regression. The crossover supports a concrete decomposition of the observed trade-off; it does not imply that input difficulty is the only mechanism in all comparisons.

\subsection{RQ4: Training Length, Failures, and Cost}
Table~\ref{tab:length} shows exact continuations rather than selecting the best seed. Mean full correctness, input kill, BTS, and full kill increase from step 50 to 75 by 1.88, 0.84, 1.52, and 0.49 points. The direction is not consistent across seeds: full kill improves only for seed 103, while seed 101 loses both correctness and utility. We therefore describe step 75 as the selected aggregate configuration on this project evaluation, not as evidence that more training is uniformly beneficial.

\begin{table}[t]
\caption{Individual Additive seeds and optimizer-preserving continuation. All values are percentages. Base is $C=28.59$, $O=24.06$, $K=12.23$, and BTS $=17.74$.}
\label{tab:length}
\centering
\begin{tabular}{rr rrrrr}
\toprule
Seed & Step & Task success & $C$ & $O$ & $K$ & BTS\\
\midrule
101 & 50 & 99.30 & 44.23 & 24.14 & 14.77 & 22.91\\
101 & 75 & 93.66 & 42.25 & 24.08 & 13.55 & 21.91\\
102 & 50 & 98.59 & 42.25 & 23.47 & 14.31 & 21.87\\
102 & 75 & 100.00 & 43.10 & 24.72 & 14.22 & 23.36\\
103 & 50 & 99.30 & 35.49 & 25.68 & 11.92 & 20.46\\
103 & 75 & 100.00 & 42.25 & 27.01 & 14.68 & 24.54\\
\bottomrule
\end{tabular}
\end{table}

Seed 101 at step 75 has nine failed tasks. Inspection of their saved responses finds an excessively long fifth input followed by an unterminated JSON response. No tests can be recovered from these model outputs, although the API requests succeed. They remain zero in the reported metrics. This example matters operationally: improved executable reward does not eliminate output-budget exhaustion, and a checkpoint can lose previously successful tasks even when its aggregate correctness remains above Base.

Recovered training logs give total wall times through step 75 of 6.01, 5.51, and 5.58 hours on four A800 GPUs. Their mean is 5.70 hours, or 22.80 GPU-hours per seed. The largest logged peak allocated memory is 51.66GB per GPU. This is framework allocation, not a measurement of total server memory or all GPU processes. Generation and executable reward handling share a timing field, so the logs do not support separating their costs after training. Saved generation latency averages 3.41 seconds per task for Base and 3.97 seconds for Additive-75, excluding offline execution and orchestration. Those observations establish the order of resource use, not a controlled serving-speed comparison.

\section{Discussion}
\subsection{The Bottleneck Is Conversion, Not Only Discovery}
Our results support separating two limits on a specification-based test generator. The first is \emph{discovery}: can it choose inputs that expose faulty behavior? The second is \emph{conversion}: can it supply correct expected outputs for those inputs? Base already has a substantial gap between potential and effective detection. Additive improves both, but the remaining gap and the highest-kill-bin failures show that finding revealing inputs does not solve the output problem.

The fixed-input experiment makes this observation more precise. FullKill's output policy is slightly better than NoKill's on identical inputs in the primary contrast, despite its lower accuracy on the inputs it chooses for itself. Thus, a decline in free-generation oracle accuracy need not mean that training has forgotten how to compute outputs. It can reflect a shift in what the policy attempts. Conversely, a rise in accuracy need not imply broader reasoning competence if the policy selects more conservative inputs. Reporting conversion together with potential kill makes both interpretations visible.

This is an empirical tension, not a mathematical incompatibility. An input can be both easy to solve and highly discriminative, and our results do not show that every higher-kill input must be harder. The policy's learned distribution and finite capacity determine how often it finds that intersection. Nor is it necessary for five tests to match the number of faulty programs. Their detection sets can overlap or cover several programs at once. Increasing suite size could increase potential coverage, but would change the inference and testing budget rather than explain the current five-test result.

\subsection{Implications for Training and Evaluation}
For a deployment whose immediate objective is effective fault detection, full kill remains the decisive testing measure. On the equal-budget comparison, NoKill is therefore a strong control, not a failed baseline to be omitted because it reduces input kill. If the research objective additionally requires preserving discriminative input generation, Additive offers a different trade-off: it avoids the mean input-kill regression while improving correctness and effective detection over Base. These requirements should be stated before interpreting a composite score.

Potential-kill reward is best viewed as a way to retain information that output errors would otherwise hide. It does not authorize using wrong-output tests, and it does not guarantee that the policy learns to repair those outputs later. The modest effective gains indicate that increasing the strength of the same input-utility signal alone is unlikely to remove the bottleneck. A targeted oracle-training intervention would need to show improvement on fixed difficult inputs and then demonstrate that the improvement survives the policy's own input choices. Our crossed design supplies those two distinct checks without requiring a second inference model.

Suite-level accounting is also necessary when exploring new objectives. Two policies can produce the same number of correct tests yet detect different wrong programs. A new correct output may add no suite utility if another test already detects every program it rejects. OCE handles this redundancy through unions, while per-input bins reveal which regions are poorly converted. Neither measure replaces direct execution of complete suites. Together they provide an explanation for a score change rather than only another score to optimize.

\subsection{Capability References}
Table~\ref{tab:references} provides context from larger models and recorded API endpoints. These are joint-generation outputs rescored on the same 142 tasks, not additional trained arms. The GPT-labeled endpoint has substantially higher full correctness and OCE than the local 9B Base, but its input kill is still 40.17\%. That observation does not identify a dataset-imposed ceiling: it is the performance of one endpoint under one prompt and budget. Its high conversion also shows that the conversion loss observed for 9B is not inevitable for every generator.

\begin{table}[t]
\caption{Capability references on the same audited task set. Values are percentages from one recorded run per model/endpoint. The API names identify saved service configurations, not independently verified model versions. These rows are not matched-compute baselines.}
\label{tab:references}
\centering\small
\begin{tabular}{lrrrrrr}
\toprule
Recorded model/endpoint & Success & $V$ & $C$ & $O$ & $K$ & OCE\\
\midrule
Qwen3.6-27B & 99.30 & 86.20 & 44.93 & 29.27 & 18.27 & 61.69\\
Qwen3.6-35B-A3B & 98.59 & 83.66 & 37.46 & 28.16 & 15.59 & 56.59\\
DeepSeek-V4-Flash & 87.32 & 75.92 & 38.73 & 25.19 & 15.33 & 60.12\\
DeepSeek-V4-Pro & 99.30 & 89.01 & 40.70 & 32.89 & 17.38 & 52.13\\
GPT-5.6-sol (proxy) & 99.30 & 84.51 & 76.20 & 40.17 & 38.61 & 95.80\\
\bottomrule
\end{tabular}
\end{table}

Differences between potential and effective rankings persist in these references. The Pro endpoint reaches more potentially detectable wrong programs than Flash, but its conversion is lower. Larger local models also retain nontrivial conversion gaps. The corresponding per-input curves are provided in Appendix~\ref{app:references}. Their bins contain different generated inputs, so cross-model comparisons cannot substitute for the fixed-input crossover or support a claim that one architecture uniquely causes the trade-off.

\section{Threats to Validity and Scope}
\label{sec:validity}
\paragraph{Adaptive use of the evaluation set.}
The 142-task set was inspected in prior experiments, and seed 103's training-length curve informed the choice of step 75 before the additional seed continuations. The three-seed results measure reproducibility under a fixed project protocol, but the selected checkpoint's absolute score is not an unbiased estimate on a newly untouched test set. We retain the step-50 comparison and every step-75 seed to make this selection history visible. The study supports an in-project result and mechanism analysis, not a broad generalization claim based solely on the selected score.

\paragraph{Reference reliability and input legality.}
Sample checks, deduplication, and dynamic execution reduce known data problems, but multiple references can share a defect or process an out-of-domain input identically. Reference-validity is consequently not a complete executable specification of input constraints. Excluding multi-output and special-judge tasks also narrows the task distribution. Our correctness claims are relative to the audited consensus protocol and deterministic-output subset, not all natural-language programming tasks. This limitation matters particularly for reward learning, because a systematic judge error can supply a consistent but misleading signal.

\paragraph{Fault-bank representativeness.}
Kill rates depend on the retained wrong programs. Exact duplicate removal avoids repeated copies inflating scores, but it does not remove every semantic redundancy or prove that each program represents a distinct fault. The small bank may omit failures that matter in deployed software, and the filtering to 5--19 wrong programs excludes other TC-Bench tasks. We evaluate finite-bank detection rather than semantic coverage of all incorrect implementations. Public problem statements may also overlap with model pretraining; the present artifacts do not permit a complete pretraining-contamination audit.

\paragraph{Optimization and mechanism inference.}
The main training comparison uses one model size and three seeds. Nested reward removals change both the utility credit and the scalar distribution entering group normalization. The crossover fixes input bytes, isolating observed source and target effects on its training-pool tasks. It cannot establish a universal relationship between input complexity and oracle difficulty. High-kill bins are sparse, and slots within a task are dependent.

\paragraph{Comparison coverage and deployment.}
The reward controls are not complete reproductions of UTGen, UTRL, CURE, or Ockhamareto. Their interfaces, supervision, and training arrangements differ, so this paper does not claim state-of-the-art performance over those systems. API references have unverified internal computation budgets and, for the proxy, unverified backend identity. Finally, we measure test correctness and program-bank detection, not a human TDD study, an end-to-end code-generation gain, or production fault prevention. Such outcomes require their own evaluation.

\section{Conclusion}
Specification-based test generation must turn revealing inputs into complete correct tests. We operationalize this requirement through potential and effective suite detection sets and oracle-conversion efficiency. A matched executable-reward study shows that correctness-oriented training improves effective kill while reducing input utility, whereas Additive GRPO preserves potential utility and yields joint improvements over Base across three step-75 seeds. The gains are modest, and Additive does not dominate the equal-budget controls. A fixed-input crossover explains the principal trade-off through harder input selection rather than uniformly degraded oracle prediction. These findings position oracle conversion as a concrete target for improving requirement-to-test generation and provide a measurement protocol for determining whether future utility gains become usable tests.

\bibliographystyle{ACM-Reference-Format}
\bibliography{references}
\appendix
\section{Prompt and Output Protocol}
\label{app:prompt}
The joint prompt is built from the following template, with the task description inserted at the end. Recognized sample sections are removed and separate sample fields are not appended. The template is shared by the matched joint-generation evaluations.
\begin{lstlisting}
You are generating complete test cases from a natural-language programming problem.

Return ONLY valid JSON, with this exact shape:
{"tests":[{"input":"...","output":"..."}]}

Rules:
- The candidate programs are C/C++ programs that read from stdin and write to stdout.
- Generate exactly 5 complete test cases. Each test case is one independent program run.
- The tests array length MUST be exactly 5. Do not output fewer tests.
- Put the exact stdin content in tests[i].input and the exact expected stdout in tests[i].output.
- Solve each generated input yourself to derive its expected output. The output must be correct.
- Use escaped newline characters in JSON strings when stdin or stdout contains multiple lines.
- Do not output Python lists, command-line arguments, checker code, or explanations.
- Do not include Markdown or explanations.
- Each input must be self-contained and valid for the problem.
- Prefer diverse tests that expose incorrect solutions: edge cases, boundary values,
  small cases whose outputs can be derived reliably, tricky structures, and representative normal cases.
- Prioritize compact counterexample-style inputs that are likely to distinguish plausible wrong
  implementations, such as off-by-one boundaries, degenerate cases, ties, overflow-sized values,
  unusual ordering, duplicate values, disconnected/empty structures, and minimum/maximum constraints.
- Avoid spending all tests on easy normal cases; include tests that would make a subtly wrong
  solution disagree with the correct output.
- Keep individual test cases compact. Do not emit very large repetitive inputs.

Problem:
{description}
\end{lstlisting}

The training wrapper applies strict JSON-schema checking before assigning its final reward. The evaluation parser instead extracts a valid JSON candidate and accepts the legacy field aliases. In both cases, malformed JSON is not repaired. The evaluator normalizes input line endings, removes empty and duplicate inputs, and retains the first five unique inputs. It preserves the generated expected outputs rather than replacing them with consensus outputs. Consensus is used only to judge correctness and construct potential detection sets.

\section{Per-Seed Reward Ablations}
\begin{table}[h]
\caption{All equal-budget ablation seeds on the fixed 142-task evaluation. Percentages use the five-slot E2E protocol. Additive seeds and continuations are in Table~\ref{tab:length}.}
\centering
\begin{tabular}{lrrrrrr}
\toprule
Method & Seed & Success & $C$ & $O$ & $K$ & BTS\\
\midrule
NoKill-50 & 101 & 100.00 & 47.18 & 22.94 & 14.38 & 23.16\\
NoKill-50 & 102 & 100.00 & 50.28 & 20.74 & 15.09 & 23.09\\
NoKill-50 & 103 & 99.30 & 43.66 & 21.11 & 13.60 & 21.95\\
FullKill-only-50 & 101 & 100.00 & 44.79 & 23.68 & 13.39 & 23.05\\
FullKill-only-50 & 102 & 97.18 & 48.87 & 23.90 & 15.75 & 25.14\\
FullKill-only-50 & 103 & 100.00 & 43.10 & 23.19 & 13.41 & 23.04\\
\bottomrule
\end{tabular}
\end{table}

\section{Cross-Model Conversion Curves}
\label{app:references}
\begin{figure}[h]
\centering
\includegraphics[width=\linewidth]{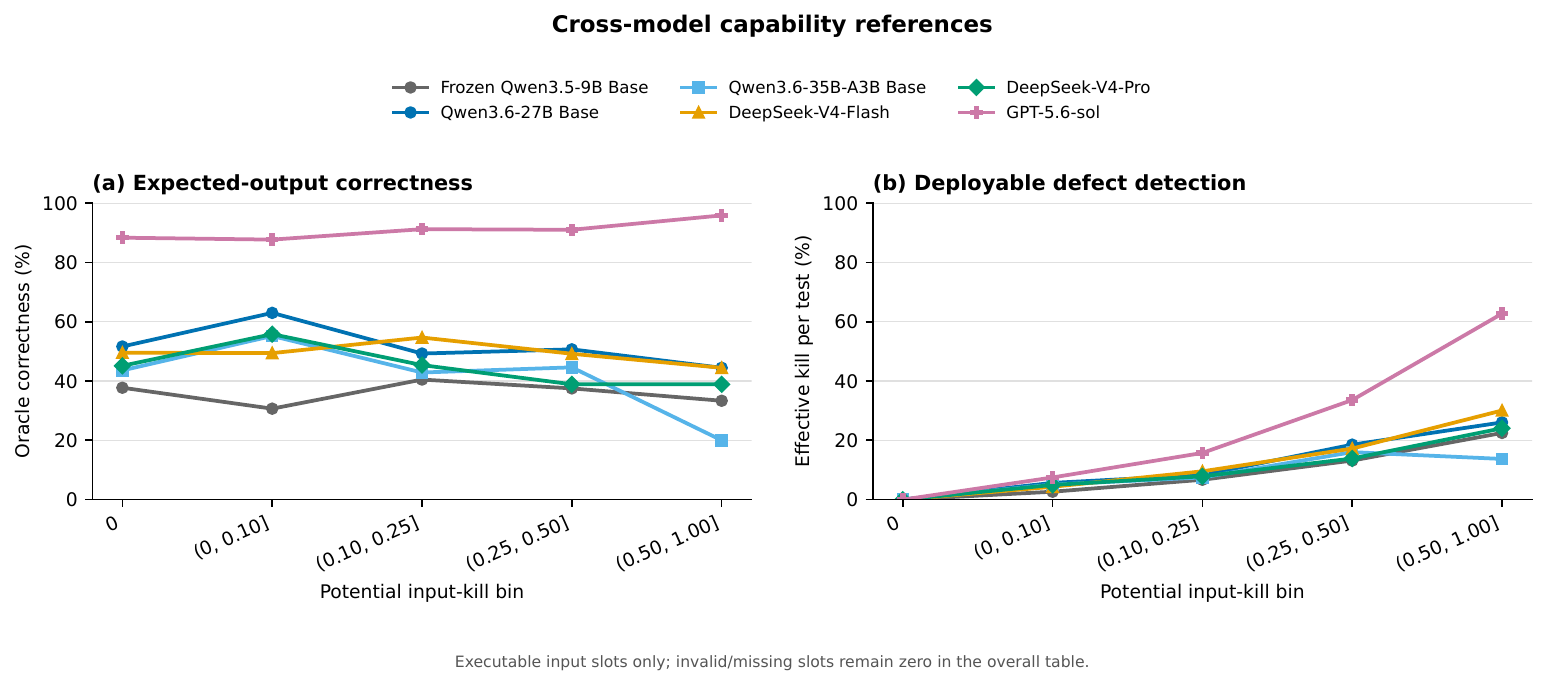}
\caption{Cross-model capability references. Curves are conditional on each model's generated reference-valid inputs. Effective per-test kill is not a suite-union score. Model labels are the recorded local or service identifiers.}
\label{fig:references}
\end{figure}

Figure~\ref{fig:references} uses the same bin definitions as Figure~\ref{fig:conversion}. The highest-kill bin contains 9 slots for Qwen3.6-27B, 10 for Qwen3.6-35B-A3B, 9 for DeepSeek-V4-Flash, 18 for DeepSeek-V4-Pro, and 24 for the GPT-labeled endpoint. These counts should accompany any interpretation of the tail. The proxy endpoint obtains correct outputs for 23 of its 24 highest-bin inputs, but those are not the inputs sampled by the 9B policies.

\section{Reproduction and Artifact Contents}
The accompanying source package contains the manuscript, BibTeX records, editable LaTeX tables, and vector figures. Every table is mapped to its source JSONL or report in the repository-side evidence manifest. The main results use Quality-v2 142-task records and seeds 101--103. Conversion analysis is offline: it reconstructs unique potential and effective program sets from saved judgments, checks their subset relation, and then aggregates them.

The training stack combines FSDP policy training and vLLM rollouts through VERL, recorded at revision \texttt{983cb0f2} in the evidence snapshot. Exact-continuation records identify the reported step-75 runs. Their logs contain 50 initial updates and 25 restored-state updates per seed, separately from the historical optimizer-reset branch.

The nine seed-101 step-75 model failures have task identifiers 22, 28, 258, 287, 434, 476, 557, 689, and 822. Their raw outputs are retained in the project records; no repaired or replacement response contributes to the reported result. The fixed-input crossover is stored separately from the 142-task evaluation, preserving its training-pool diagnostic role. Program sources, model weights, API credentials, and user-specific filesystem paths are not embedded in the manuscript source archive.

\end{document}